\documentclass{osa-article}

\journal{boe}

\articletype{Research Article}
\begin{document}

\title{Initial non-invasive in vivo sensing of the lung using time domain diffuse optics}

\author{Antonio Pifferi,\authormark{1,2} Massimo Miniati,\authormark{3}  Andrea Farina,\authormark{2,*} Sanathana Konugolu Venkata Sekar,\authormark{4} Pranav Lanka,\authormark{1} Alberto Dalla Mora,\authormark{1} and Paola Taroni\authormark{1,2}}

\address{\authormark{1}Politecnico di Milano, Dipartimento di Fisica, Piazza Leonardo da Vinci 32, Milano, Italy\\
\authormark{2}INFN-CNR, Consiglio Nazionale delle Ricerche, Istituto di Fotonica e Nanotecnologie, Piazza Leonardo da Vinci 32, Milano, Italy\\
\authormark{3}University of Florence, Department of Experimental and Clinical Medicine, Italy\\
\authormark{4}Biophotonics@Tyndall, IPIC, Tyndall National Institute, Lee Maltings, Dyke Parade, Cork, Ireland}

\email{\authormark{*}andrea.farina@cnr.it} 



\begin{abstract*}
Non-invasive \textit{in vivo} sensing of the lung with light would help diagnose and monitor pulmonary disorders (caused by e.g. COVID-19, emphysema, immature lung tissue in infants). We investigated the possibility to probe the lung with time domain diffuse optics, taking advantage of the increased depth (few cm) reached by photons detected after a long (few ns) propagation time. An initial study on 5 healthy volunteers included time-resolved broadband diffuse optical spectroscopy measurements at 3~cm source-detector distance over the 600-1100~nm range, and long-distance (6-9~cm) measurements at 820~nm performed during a breathing protocol. The interpretation of the \textit{in vivo} data with a simplified homogeneous model yielded a maximum probing depth of 2.6-3.9~cm, suitable to reach the lung. Also, signal changes related to the inspiration act were observed, especially at high photon propagation times. Yet, intra- and inter-subject variability and inconsistencies, possibly alluring to competing scattering and absorption effects, prevented a simple interpretation. Aspects to be further investigated to gain a deeper insight are  discussed.\end{abstract*}


\section{Introduction\label{SecIntro}}
The lungs can be regarded as a mixture of air and tissue with a water-equivalent density. Accordingly, they can be visualised noninvasively through the attenuation of x-rays from an external source. In clinical practice, both conventional chest radiography (CXR) and computed tomography (CT) are widely used to detect pulmonary abnormalities. Thanks to its inherent tomographic properties, CT is superior to conventional CXR in visualising the fine structure of the lung. Modern CT scanners are capable of detecting small solid lesions down to a size of 5~mm in width. However, neither CXR nor CT provide any direct information as regards the chemical composition of an abnormal lung density. The same limitation applies to thoracic ultrasound (US) which is increasingly used by the clinicians to evaluate pleural effusions or subpleural consolidations at bedside.

Assessing the lung composition and density can be useful to diagnose or monitor patients with emphysema, a pathologic condition characterized by an increased fraction of air per unit of lung volume. In patients with left heart failure, lung density may increase substantially due to the accumulation of water in the extravascular space of the lung (pulmonary edema). Also, lung inflammations with fluid accumulation (caused e.g. by COVID-19 or other infections) leads to changes in lung density and composition. Lung density can be equally increased by an abnormal deposition of collagen in the lung interstitium (fibrosis).

Time-Domain Diffuse Optical Spectroscopy (TD-DOS) is a photonics technique that permits to derive information on the characteristics of a biological tissue through the absorption ($\mu_a$) and reduced scattering ($\mu_s'$) coefficient via the study of photon propagation of short (ps) laser pulses within the medium\cite{Durduran2010,Pifferi2016NewReview}. So far, TD-DOS had been successfully applied to characterise nonivasively the tissue composition or function of organs, like the brain\cite{Torricelli2014TimeMapping.} or the breast\cite{Grosenick2016ReviewSpectroscopy}, which are in close proximity to the surface of the body or can be assessed in transmittance geometry.

From the optical standpoint, the lungs are accessible over a large portion of the anterior and lateral chest wall. However, the varying thickness of the chest wall and its heterogeneous composition (collagen, elastin, fat, and muscular tissue) pose a major challenge when using TD-DOS transcutaneously. In addition, the varying dimensions of the alveoli and distal conducting airways over the respiratory cycle may significantly affect the scattering of photons, making the interpretation of experimental data specially complex\cite{Durkee2018LightModelb}.

As of now, optical techniques have been applied mostly for the diagnosis of lung cancer through invasive (bronchoscopy), or minimally invasive (needle biopsy) techniques\cite{Zeng2004OpticalReview}. They include diffuse reflectance spectroscopy\cite{Spliethoff2013ImprovedSpectroscopy,Spliethoff2016Real-timeStudy}, endogenous\cite{Lam2000DetectionBronchoscopy} or exogenous fluorescence spectroscopy\cite{Okusanya2015IntraoperativeResection,Hernot2019LatestSurgery}, optical coherent tomography \cite{Lam2008InLesionsb}, and Raman spectroscopy \cite{McGregor2017Real-timeDetection}. Much effort is also being devoted to model the photon propagation through the lung by means of simulations\cite{Sikorski2006ModelingSolver,Singh2015NanoparticleRespirationb} anthropomorphic lung phantoms\cite{Larsson2018DevelopmentConsiderations,Pacheco2020AnthropomorphicInfants}, and \textit{ex vivo} bovine lung tissues\cite{Lacerenza2020FunctionalSample}.
Conversely, there are not many attempts to probe the lung optically non-invasively transcutaneously. In a first clinical study on 3 full-term babies, an optical technique termed "Gas in Scattering Media Absorption spectroscopy" (GASMAS) was used for noninvasive monitoring of water vapor in the lungs\cite{Lundin2013NoninvasiveStudy}. A larger study on 29 newborn infants demonstrated transcutaneous detection of oxygen in the lungs using GASMAS\cite{Svanberg2016DiodeInfants}. This approach was further validated also on a piglet animal model\cite{Svanberg2020ChangesPiglets}. Yet, to the best of our knowledge, so far no non-invasive optical approach has been applied \textit{in vivo} on the lungs in adults.

In the present study, we investigated the feasibility to sense the lung transcutaneously by simulations – instructed by \textit{in vivo} spectroscopy measurements on the thorax – and by applying TD-DOS in healthy volunteers who were examined at suspended full inspiration and suspended full expiration.

\section{Materials and methods\label{SecMatMet}}
\subsection{System set-up}
Two TD-DOS systems were used for \textit{in vivo} measurements on volunteers, namely a broadband spectrometer and a highly sensitive single-wavelength setup.

The broadband spectrometer was a laboratory workstation exploiting supercontinuum generation to automatically perform time-domain measurements of the Distribution of Time-of-Flight (DTOF) over a wide spectral range, from 500 up to 1700 nm encompassing different detectors\cite{Sekar2017DiffuseNm}. In the actual embodiment, an 80~MHz, 10~ps pulsed supercontinuum (SuperK EXTREME, NKT photonics) was sliced using a rotating Pellin-Broca prism followed by a 50~$\mu m$-core graded index fiber acting as spectral selection. Light was injected to and collected from the tissue using 1~mm core step index fibers coupled to a custom-made Silicon Photomultiplier (SiPM) module\cite{Martinenghi2015SpectrallySpectroscopy} connected to a Time-Correlated Single-Photon Counting (TCSPC) board (SPC130, Becker \& Hickl). A variable optical attenuator set on the source path controlled photon counting rate within the 1\% single-photon statistics. A complete spectral acquisition from 600 nm to 1100 nm in steps of 10 nm with 4 s acquisitions required around 5 min.

The single-wavelength setup was optimised for maximal light harvesting. It was based on a prototypal 40 MHz, 25 ps high-power Four-Wave Mixing laser (Fianium Ltd, Southampton, UK). The tissue was illuminated at 820 nm by a maximum power of 100 mW expanded to comply with the maximum permissible skin exposure. Re-emitted light was harvested using 1-mm core step index fiber coupled to a hybrid photomultiplier (HPM-100-50, Becker \& Hickl GmbH, Germany), and connected with a TCSPC board (SPC130, Becker \& Hickl GmbH, Germany).

\subsection{Simulations}
To simulate the DTOFs in a layered tissue, a time-resolved GPU-accelerated Monte Carlo code \cite{Alerstam2008ParallelMigration} was implemented making use of the microscopic Lambert-Beer approach \cite{Sassaroli2012EquivalenceMedia}: for each detected photon the path-length spent in each layer was saved. The simulation was run without absorption, weighting each trajectory afterwards depending on the desired absorption in each layer. A total of 10M photons reaching the receiver were collected for each simulation.

\subsection{In vivo protocol}
Five healthy volunteers were recruited for the study, after signing informed consent and complying with the authorization from the institutional Ethical Review Board of Politecnico di Milano. Demographic data are reported in Table~\ref{TabComp} presented in Section~\ref{SecSpectra}.

For the broadband study, the subject was lying supine breathing normally and with the probe positioned on the thorax in the upper-right region using a source-detector distance $\rho=3$ cm.

For the single-wavelength study, the subject was lying supine, with the probe set in two locations on the thorax, namely on the anterior surface of the right hemithorax between the second (UR) and fifth (DR) intercostal space along  the mid clavicular line. The source-detector distance $\rho$ was set to the maximum value ensuring a count-rate around 1 M counts/s, resulting in $\rho=6-9$ cm depending on the subject and location. The subjects were asked to follow a suspended full inspiration and suspended full expiration at a given pace to provide two intervals with the lung inflated (IN) or deflated (OUT). Two protocols were applied on each subject with 5 repetitions of the two phases lasting 10 s each (Prot10) \textit{or} 10 repetitions with 5 s for each phase (Prot5). Only for Prot5 the first repetition was perfored at normal breathing to get a reference state. Each protocol was repeated twice in each location, resulting in 8 complete series of acquisitions per subject. 

\subsection{Data analysis}
The spectral measurements were analysed using a homogeneous solution of the Diffusion Equation under the extrapolated boundary conditions using $\mu_a$ and $\mu_s'$ as free fitting parameters. The Instrument Response Function (IRF) obtained facing the injection and collection fibers was convoluted with the theoretical model. The fitting range included points with a number of counts >80\% and >1\% of the peak value on the leading and falling edge of the DTOF, respectively. Tissue composition was derived from the retrieved absorption and reduced scattering spectra using the linear combination of 5 key tissue absorbers, namely oxy- (0$_2$Hb), and deoxy-hemoglobin (Hb), water, lipids, and collagen.

The single-wavelength measurements were analysed using a fit with the homogeneous model similarly to the spectral measurements. For better visualization, a folding average over the 5 (10) repetitions of the protocol was applied. Further, the gated intensity was calculated for different time-windows of the DTOFs. The relative contrast of the gated reflectance signal $R(t)$ with respect to a reference state $R_0(t)$ was derived as:

\begin{equation} \label{EqContrast}
C(t)=\frac{\int_{t}^{t+\Delta t} R(t')dt' - \int_{t}^{t+\Delta t} R_0(t')dt'}{\int_{t}^{t+\Delta t} R_0(t')dt'} \end{equation}

where $t'$ is the photon arrival time, while $t$ and $\Delta t$ are the starting edge and the width of the gated temporal window. For the \textit{in vivo} protocols, $R_0$ is assumed to be the average of $R$ over the whole exercise, therefore $C$ represents the relative change in gated signal with respect to the mean value.

Equation~\ref{EqContrast} was used also to calculate the relative contrast for the Monte Carlo simulations. In that case, for $R_0$ we assumed the unperturbed state. 

\section{Spectroscopy of the human chest\label{SecSpectra}}
\begin{figure}[ht!]
\centering\includegraphics[width=12.5cm, height=5cm]{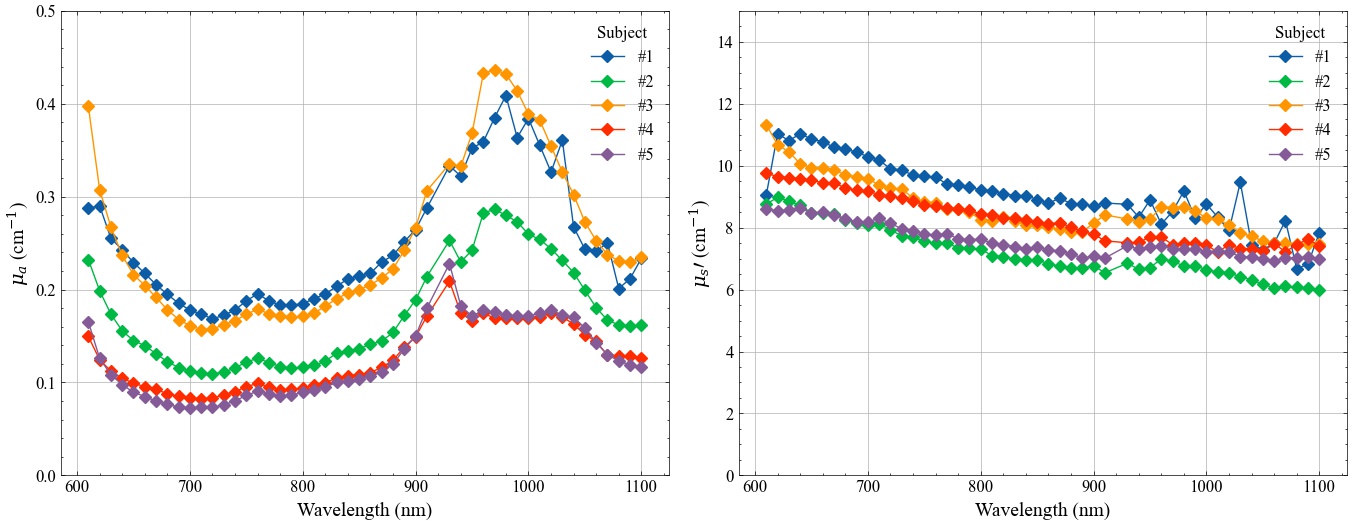}
\caption{Absorption (left) and reduced scattering (right) spectra on the chest of 5 healthy volunteers using a source-detector distance $\rho = 3$ cm and a homogeneous model.}
\label{FigSpectra}
\end{figure}

As a first step to estimate the optical properties of the tissues overlaying the lung, we measured the absorption and reduced scattering spectra on the chest region of healthy volunteers using an intermediate source-detector distance $\rho=3$ cm. Figure~\ref{FigSpectra} displays the absorption (left pane) and reduced scattering spectra (right pane) of 5 subjects (see Table 1 for demographics) obtained assuming a homogeneous model. Some inter-subject variability in the absorption spectrum is observed, with a pronounced lipid contribution for subject \#5, and a high muscular presence for subject \#1 and \#3. The clear peak around 980 nm is ascribed to water, while the lipid peak at 930 is progressively covered by the dominant water contribution, when moving from \#5 to \#1. Differences in the region <800 nm are mostly due to blood content. Overall, around 800 nm – where we performed large-$\rho$ measurements as presented in the following Section~\ref{FigSim} – $\mu_a \approx 0.1-0.2$ cm$^{-1}$ and $\mu_s' \approx 7-9$ cm$^{-1}$. Table~\ref{TabComp} reports the average tissue composition assuming 5 absorbers (Hb, HbO$_2$, water, lipids, and collagen) and an empirical power law dependence for the scattering spectrum choosing $\lambda_0=600$ nm\cite{Mourant1997PredictionsPhantoms}:
\begin{equation} \label{EqMie}
\lambda=a(\lambda/\lambda_0)^{-b}
\end{equation}
The mean tissue composition of different subjects confirms the initial guess on the fat -- or muscle -- predominance.

\begin{figure}[ht!]
\centering\includegraphics[width=9cm]{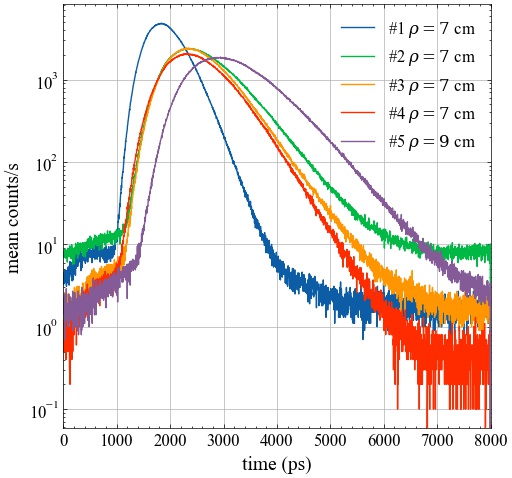}
\caption{Distribution of photon times of flight (DTOFs) for the 5 volunteers on the chest in reflectance geometry using the largest source-detector distance ($\rho=7 \sim 9$ cm) yielding a full count-rate collection statistics of $\approx 1$ Mcounts/s. The maximum photon travelling time is up to $4 \sim 8$ ns depending on the subject.}
\label{FigDTOF}
\end{figure}

To gain an initial guess on the maximum detectable photon traveling time, which determines the maximum explored depth in the medium\cite{Torricelli2005Time-ResolvedImaging,Martelli2016TheresMediab}, in Fig.~\ref{FigDTOF} we show the DTOFs measured on the chest of the same 5 subjects using quite a large $\rho=7-9$ cm. We adopted here the single-wavelength system while choosing the maximum $\rho$ permitting to reach the max count rate for the instrumentation (1 M counts/s). Increasing $\rho$ beyond that limit would simply decrease the number of collected photons at any times\cite{Torricelli2005Time-ResolvedImaging}. As expected, the temporal profile is quite broad and extends to late times, with early photons arriving not earlier than 1 ns after the pulse injection time ($t=0$ ns). The maximum photon arrival time, or better the max $t$ when the DTOF stands out of the noise, ranges from 4 ns for subject \#1 up to 8 ns for subject \#5. These values can provide a first hint to interpret the simulations presented below, and to guess the maximum investigation depth of the measurements. The absorption and scattering properties derived using a homogeneous model for this large $\rho$ at 820 nm are also reported in Table~\ref{TabComp}.

\begin{table}[ht!]
\centering
\caption{Subjects' demographics, mean composition, and optical properties at 820 nm.}
\label{TabComp}
\begin{tabular}{c|c|c|c|c|c|c|c|c|c|c|c} 
\hline
subject & age & BMI & Hb & HbO$_2$ & water & lipid & collagen & a & b & $\mu_a$(*) & $\mu_s'$(*)\\
& (y) & (Kg/m$^2$) & $\mu M$ & $\mu M$ & g/cm$^3$ & g/cm$^3$ & g/cm$^3$ & cm$^{-1}$ & & cm$^{-1}$ & cm$^{-1}$\\
\hline
\hline
\#1 &  &  & 1.5 & 25.1 & 0.34 & 0.43 & 0.21 & 11.1 & 0.63 & 0.32 & 11.7\\
\hline
\#2 & 65 & 23.1 & 3.0 & 15.5 & 0.29 & 0.63 & 0.11 & 9.0 & 0.78 & 0.14 & 7.3\\
\hline
\#3 & 52 & 19.1 & 6.9 & 26.4 & 0.49 & 0.67 & 0.14 & 10.5 & 0.75 & 0.34 & 11.6\\
\hline
\#4 & 34 & 24.1 & 0.8 & 11.4 & 0.12 & 0.69 & 0.04 & 10.0 & 0.59 & 0.15 & 8.2\\
\hline
\#5 & 49 & 25.8 & 1.3 & 15.8 & 0.14 & 0.85 & 0.04 & 9.0 & 0.61 & 0.09 & 5.3\\
\hline
\hline
(*)820nm
\end{tabular}
\end{table}

\section{Simulations\label{SecSim}}
\begin{figure}[ht!]
\centering\includegraphics[width=9cm]{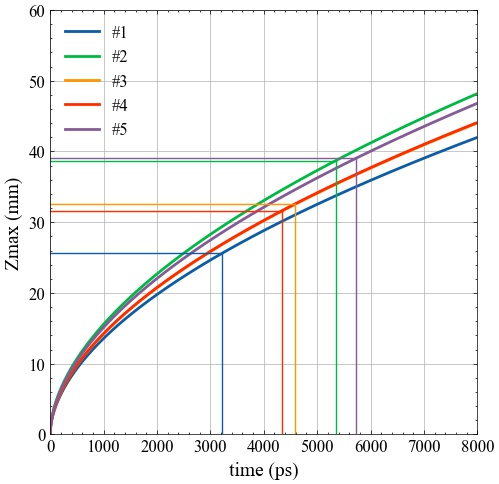}
\caption{Mean value of the maximum depth of photon trajectories ($z_{max}$) as a function of photon arrival time derived for the $\mu_s'$ properties of the 5 subjects. The straight lines indicate the maximum $t$ yielding a count-rate >10,000 counts/s over a late gate for data in Fig.~\ref{FigDTOF}, and the corresponding value of $z_{max}$}
\label{FigZmax}
\end{figure}

As an initial step to answer the question whether it is possible to reach the lung, we addressed the problem from a theoretical point of view, studying the maximum depth reached by photons in a homogeneous diffusive medium in reflectance geometry. Here, we adopted the approach proposed in Ref.\cite{Martelli2016TheresMediab} that derives the mean value of the maximum depth $z_{max}$ of all possible photon trajectories in a semi-infinite medium (see Material and Methods for details). Figure~\ref{FigZmax} displays $z_{max}$ as a function of photon travelling time $t$ for the $\mu_s'$ derived for each subject at 820 nm with $\rho = 3$ cm (data from Fig.~\ref{FigSpectra}), which reasonably represents the mean scattering properties in the superficial chest layers than need to be traversed to reach the lung. For a homogeneous medium, the mean depth is rigorously independent of both $\mu_a$ and $\rho$, which affect the number of collected photons but do not alter the depth distribution of photon trajectories for a given $t$. Calculating the maximum photon arrival time $t_{max}$ yielding >10,000 counts/s for a late gate (shot noise = 1\%) we derive $t_{max} \approx 3.2-5.7$ ns, which leads to $z_{max} \approx 2.6-3.9$ cm for different subjects, as represented by lines in Fig.~\ref{FigZmax}.

\begin{figure}[ht!]
\centering\includegraphics[width=12.5cm]{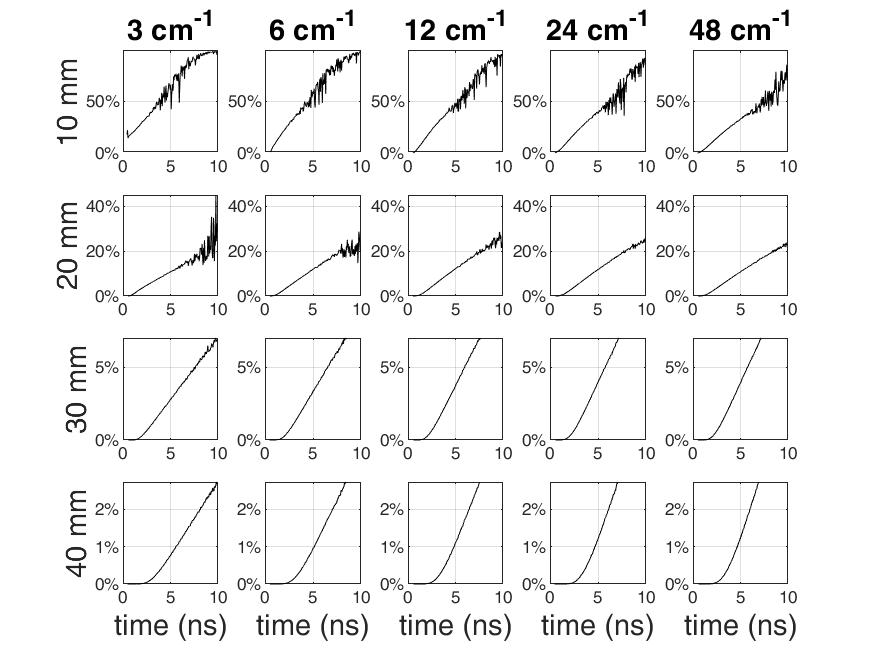}
\caption{Monte Carlo simulations of the relative contrast $C$ as a function of photon arrival time $t$ for an absorption reduction of 10\% in the lung absorption for different values of $\mu_s'$ assumed homogeneous (columns) and of the lung depth (rows).}
\label{FigSim}
\end{figure}

As a more realistic scenario, we assumed a two-layer medium composed of the lung (bottom layer) and all the overlaying tissues (top layer). Using a Monte Carlo code, we simulated a 10\% reduction in the lung absorption. Figure~\ref{FigSim} shows the relative contrast $C$ between the perturbed and unperturbed states plotted as a function of the photon arrival time $t$ for different depths of the lung layer $z_{lung}$ (rows) and for different choices of $\mu_s'$ in the lung (columns), while assuming $\mu_s'=7$~cm$^{-1}$ for the upper layer. A relative contrast $C \approx 2-3\%$ is achieved down to $z_{lung} \approx 3$~cm for $t=4$~ns, and even down to $z_{lung} \approx 4$ cm for $t=8$~ns. Quite surprisingly, the lung scattering does not affect much $C$. Yet this is in agreement with the previous argument that the photon travelling depth is mostly affected by the upper layer. This result relaxes the threat that an extremely high $\mu_s'$ in the lung could completely prevent lung sensing. Indeed, there is not yet a sound knowledge in literature of the lung scattering properties, and we opted to span a wide range. The scattering of the upper layer could be an additional variable, although we observed limited inter-subject variability (see Fig.~\ref{FigSpectra} and Table~\ref{TabComp}).

\section{\textit{In vivo} breathing protocol\label{SecBreath}}
As an \textit{in vivo} trial of the actual chances to sense the lung, we performed \textit{in vivo} measurements on the 5 subjects following a breathing protocol as described in Section~\ref{SecMatMet}. The aim of this protocol was to produce a decrease in both $\mu_a$ and $\mu_s'$ of the lung during the inhalation phase, due to the reduction of lung density, and check the sensitivity of the optical measurement to such changes.

\begin{figure}[ht!]
\centering\includegraphics[width=12.5cm, height=5cm]{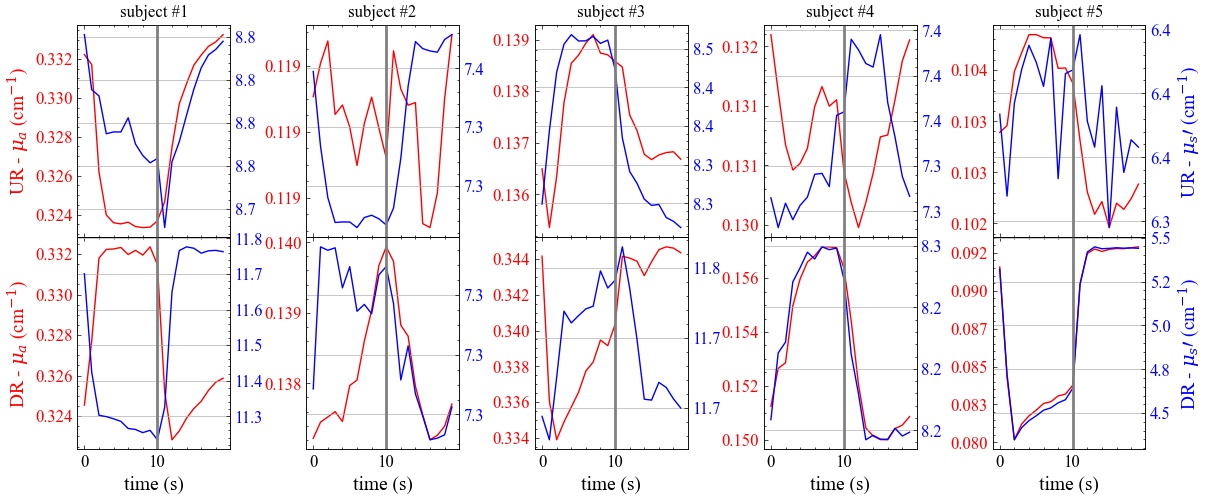}
\caption{\textit{In vivo} time evolution of the absorption (red) and reduced scattering (blue) coefficients at 820 nm during the inhalation protocol Prot10 for the 5 volunteers (rows) and 2 locations (columns) over the lung. At $t=0$ s, the subject is asked to inhale, while at $t=10$ s the subject starts to exhale.}
\label{FigOpt}
\end{figure}

Figure~\ref{FigOpt} reports the evolution of $\mu_a$ (red, left axis) and $\mu_s'$ (blue, right axis) as obtained using a homogeneous fit over the time of the task for the 5 subjects (columns) and 2 positions on the chest, namely an up-right location (UR, top row) and a down-right location (DR, bottom row). The protocol in use was Prot10 and the folding average over the exercise time was applied. In general, there is a task-related change in the fitted optical properties. Yet, the results are somehow contradictory with large inter- and intra-subject differences. In some cases, (e.g. \#5-DR), both $\mu_a$ and $\mu_s'$ decrease with inhalation as expected from the reduction in lung density, but in other cases the behavior is just the opposite (e.g. \#3-UR), or even a parameter increases while the other decreases (e.g. \#1-DR). Also intra-subject differences among the two positions are observed (e.g. \#1). The task-related alteration is observed also in the not-refolded time evolution (see Supplementary Fig.~S\ref{FigOpt_b}), yet again with quite different patterns. According to our own experience in TD-DOS, a homogeneous fit of a large-$\rho$ measurement typically enhances the optical properties of the lower (few cm depth) layer\cite{Pifferi2001ReconstructionSpectroscopyb}. This is particularly true for $\mu_a$, while information on $\mu_s'$ is more superficial\cite{Pifferi2001ReconstructionSpectroscopyb}. The other protocol (Prot5) shows substantially similar results (see Supplementary Fig.~S\ref{FigOpt_c} and Fig.~S\ref{FigOpt_d} for the re-folded and not-folded representation, respectively).

\begin{figure}[ht!]
\centering\includegraphics[width=12.5cm, height=5cm]{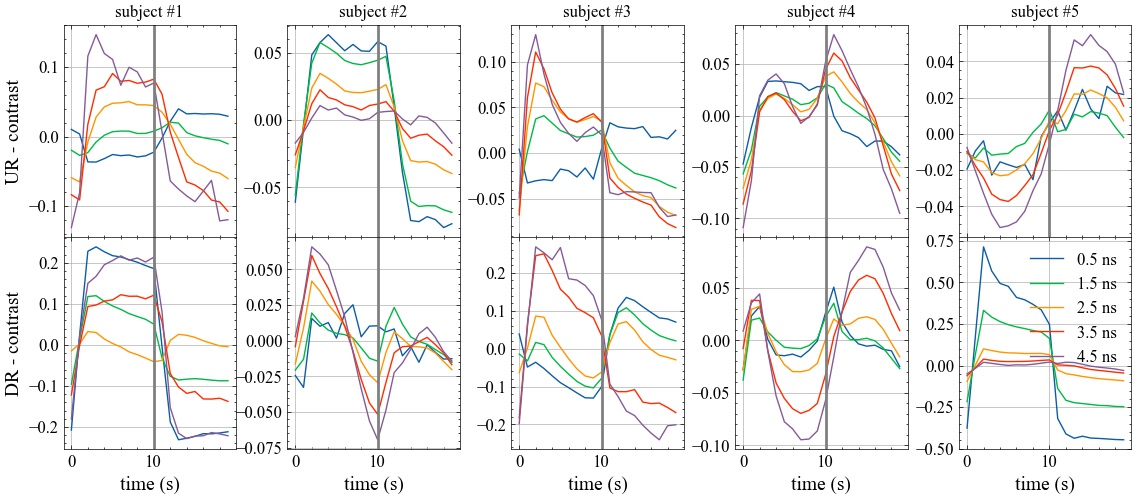}
\caption{\textit{In vivo} time evolution of the relative contrast for different time gates (see legend) at 820 nm during the breathing protocol Prot10 for the 5 volunteers (columns) and 2 locations (rows) over the lung. At $t=0$ s, the subject is asked to inhale, while at $t=10$ s the subject starts to exhale.}
\label{FigGate}
\end{figure}

As a different view to the same data, we analyzed the relative contrast $C$ in photon counts for different time gates over the IN and OUT phases, respectively. Figure~\ref{FigGate} displays $C(t)$ for 5 different gates, with $t$ ranging from 0.5 to 4.5 ns and $\Delta t=0.5$ ns as a function of the refolded time for the 5 subjects (columns) and the 2 positions (rows). The Prot10 protocol with folding average is displayed here. In all 10 plots, apart from \#5-UR and possibly \#4-DR, we observe a reduction in $C$ for late gates, that is a reduction in signal intensity when the lung tissue gets denser (OUT phase). The other protocol Prot5 shows substantially similar results (see Supplementary Fig.~S\ref{FigGate_b} for the not-folded Prot10, and Fig.~S\ref{FigGate_c} and Fig.~S\ref{FigGate_d} for the Prot5 in re-folded and not-folded visualization, respectively).

\begin{figure}[ht!]
\centering\includegraphics[width=12.5cm, height=5cm]{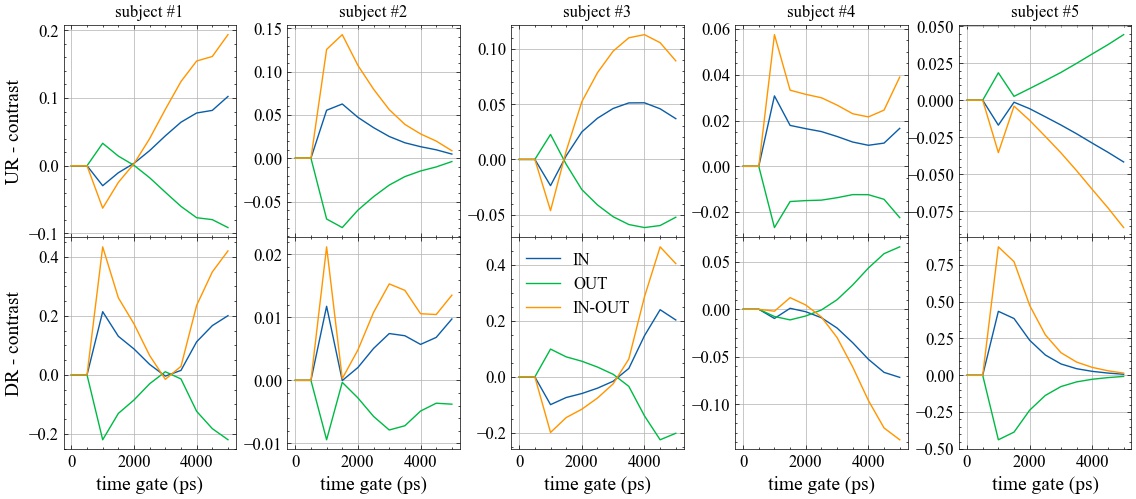}
\caption{Dependence of the relative contrast $C$ $vs$ photon propagation time during the plateau of the inhale (IN) or exhale (OUT) phases, together with the difference (IN-OUT) for the 5 volunteers (columns) and 2 locations (rows). In most cases, $C$ (i.e. the collected signal) increases from the OUT to the IN phase, yet with strong inter- and intra-subject variability and complex evolution for increasing photon propagation time.}
\label{FigMean}
\end{figure}

In order to better visualize the signal perturbation as a function of the photon arrival time $t$, we calculated the average contrast with respect to the plateau state of the inspiration and expiration phases (breath hold). Figure~\ref{FigMean} shows the mean value of $C(t)$ during the IN phase (blue line), the OUT phase (green line) and the IN-OUT difference (orange line) as a function of the gate time $t$. As for Fig.~\ref{FigOpt} and Fig.~\ref{FigGate}, columns correspond to the subject, while rows to the position. In general terms, the IN-OUT line is positive for late gates for most subjects as expected. Yet, following simulations in Fig.~\ref{FigSim}, $C(t)$ should be monotonically increasing with $t$ due to the increase in mean probed depth and therefore in lung contribution. The situation depicted in Fig.~\ref{FigMean} is more complex. There seem to be two opposite competing factors causing the IN-OUT curve to cross the null contrast at a given $\bar{t}$ and then to diverge for larger $t$ values. Similar results are observed for the Prot5 protocol (see Supplementary Fig.~S\ref{FigMean_b}).

All these \textit{in vivo} results seem to point out that we are possibly reaching the lung, but there are competing factors, making the interpretation of the breathing protocol not straightforward. In the following, we analyse five issues to be considered.

\begin{figure}[ht!]
\centering\includegraphics[width=8cm, height=8cm]{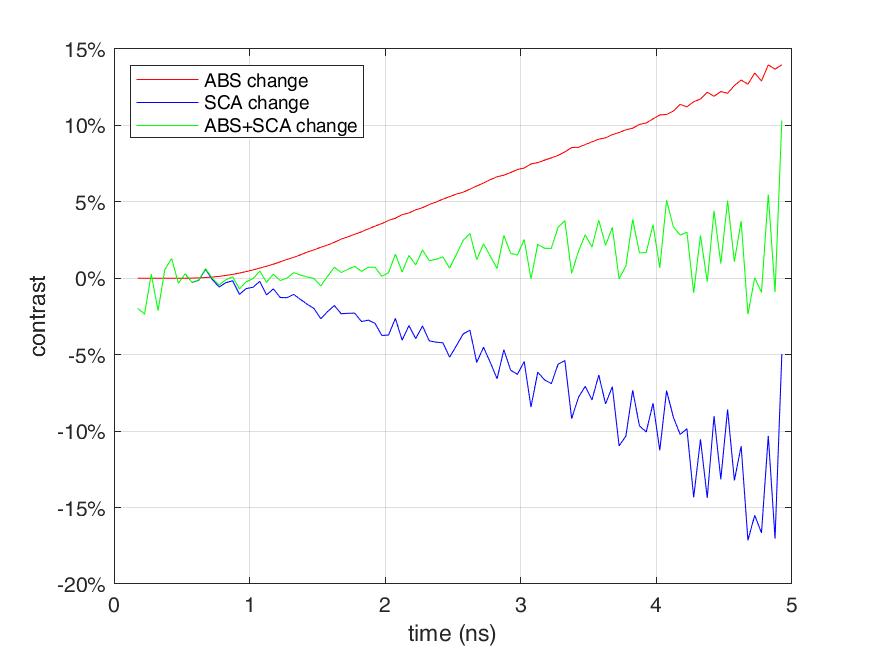}
\caption{Monte Carlo simulation of the relative contrast as a function of the photon arrival time ($t$) assuming an increase in lung air content leading to a 20\% reduction only in $\mu_a$ (ABS) in $\mu_s'$ (SCA) or rather the combined reduction of both (ABS+SCA).}
\label{FigOpposite}
\end{figure}

The first factor is the interplay of $\mu_a$ and $\mu_s'$ in the lung during inhalation. Both of them should decrease, due to the lower tissue density, yet the forecast of scattering reduction is not necessarily linear with tissue density due to the contribution of air-filled alveoli as independent scattering centers\cite{Beek1997TheRespiration}. Even more intriguing, complex phenomena such as anomalous diffusion due to the air gaps could be invoked\cite{Barthelemy2008ALight}. As a very simple, extreme case, we plot in Fig.~\ref{FigOpposite} the Monte Carlo simulation of the contrast $C(t)$ assuming an identical reduction by 20\% of $\mu_a$ and $\mu_s'$ in the lung. A reduction \textit{only} in absorption (red line) yields a linear increase of $C$ \textit{vs} $t$. A reduction \textit{only} in scattering (blue line) yields a linear decrease of $C$ \textit{vs} $t$ of a comparable amount. The combined effect of both parameters (green line) dramatically reduces the overall contrast.  

A second troubling factor resides in the lung scattering, which could be extremely high, thus preventing its exploration. There is still poor knowledge on the lung optical properties, particularly \textit{in vivo}. Some \textit{ex vivo} studies report different values for $\mu_s'$ reaching up to $20$ cm$^{-1}$ at 800 nm (calculated using Eq.~\ref{EqMie} and tabulated $a$ and $b$ values)\cite{Jacques2013OpticalReview}, but this high value could be due to the room temperature adopted in measurements leading to overestimation of $\mu_s'$ with respect to the \textit{in vivo} environment. At any rate, simulations presented in Fig.~\ref{FigSim} indicate that the scattering of the lower (lung) layer is not so critical in determining the sensitivity to lung properties beyond a given photon travelling time over a large range of lung $\mu_s'$ values.

A third factor to be considered is the pleura, which is the two-foil membrane enveloping the lung and the chest wall, separated by a lubrication fluid, which permits smooth sliding and expansion of the lung during the inhalation act. From the optical point of view, the pleura could cause a light-guiding effect, creating contamination from more superficial photons even at a large $t$ values, thus covering deep photon signals. Also in the brain, the cerebrospinal fluid (CSF) was questioned to cause light-guiding effects in functional Near Infrared Spectroscopy measurements (fNIRS). This issue was long discussed in the initial ages of fNIRS with opposite conclusions, but now the question is definitely settled by strong evidence of cortex-related activation in many studies and agreement with fMRI scans\cite{Eggebrecht2014}. The pleura thickness - extrapolated from animal data - could be in the $30-40 \mu m$ range\cite{Lai-Fook2004PleuralExchange}, so definitely lower than the CBF layer.

A forth aspect is the inadequacy of the simple analysis tools used in this study to describe the complex chest structure. The homogeneous fit proposed in Fig.~\ref{FigOpt} is a clear oversimplification of the actual geometry. Yet, in previous works confirmed with phantom experiments, we demonstrated that $\mu_a$ spectra obtained for reasonable large $t$ values using a homogeneous fit tend to adhere to the lower (few cm deep) layer\cite{Pifferi2001ReconstructionSpectroscopyb}. The photon pathlength in the upper layer reaches quickly a stable plateau after a given $t$, while the pathlength in the lower layer linearly increases with $t$ above the same threshold. Thus, the fit on $\mu_a$ is mainly affected by the lower layer properties being related to the temporal slope on the tail of the DTOF. Still, this simple argument does not hold true for $\mu_s'$, and in our case the variation in both $\mu_a$ and $\mu_s'$ might not be properly captured by this simple model.

A fifth concern is related to the changes in optical properties in the lung during the inhalation phase and the conversion of Hb to O$_2$Hb. Still, the adopted wavelength (820 nm) is close to the isosbestic point and should be weakly dependent on the oxygenation status. Also, in the long breath holding protocol (Prot10) the observed changes are quite rapid and possibly related to the inhalation act and not continuously increasing with laboratory time as if the reason were the progressive uptake of O$_2$.

\section{Conclusion\label{SecConc}}
We have presented the first study on the transdermal optical accessibility of the lung using time domain diffuse optics based on Monte Carlo simulations and \textit{in vivo} measurements on 5 healthy volunteers.
\textit{In vivo} broadband absorption and reduced scattering spectra in the 600-1100 nm range at $\rho=3$ cm analyzed using a homogeneous model permitted to estimate a possible range of optical properties mostly ascribed to the tissues to be traversed to reach the lung. In particular, around 800 nm we obtained $\mu_a=0.1-0.2$ cm$^{-1}$, and $\mu_s'=7-9$ cm$^{-1}$. Even at a larger $\rho$ ($7-9$ cm), the measured optical properties at 820 nm fall in a similar range. The same large-$\rho$ \textit{in vivo} measurements showed that a photon travelling time up to $t \approx 3.2-5.7$ ns with at least 10,000 counts/s (shot noise=1\%) in a late gate could be reached for different subjects.
Also, two-layer Monte Carlo simulations with different choices of lung $\mu_s'$ in a wide range of values ($3-48$ cm$^{-1}$) showed a relative contrast $C \approx 2-3\%$ for a 10\% reduction in lung $\mu_a$ down to a depth of $3$ $(4)$ cm for $t=4$ $(8)$ ns, which is well beyond the average lung depth.
The \textit{in vivo} measurements on a paced breathing protocol showed clear task-related changes in time domain signals. Yet, the analysis of DTOFs using a homogeneous model showed contradictory results, not in agreement with the expected decrease in both $\mu_a$ and $\mu_s'$ during inhalation. Conversely, the plot of the relative contrast $C$ for the photon counts in temporal gates showed a more consistent trend with a general increase in reflectance signal during the inhalation phase, as expected from diffuse optics models. Still, the contrast is not always increasing upon increasing photon arrival times, as foreseen by a two-layered model, and there seems to be competition between opposite factors.

In conclusion, we do not have yet a definitive sound evidence to be able to detect lung-related properties \textit{in vivo}. Yet, we contribute to the knowledge base on this intriguing quest by providing optical properties from the chest, clear simulation scenarios and a potentially powerful protocol for \textit{in vivo} validation. We plan to deploy the whole dataset in an open data paper for future analysis with more refined models. Technology is advancing at a fast pace in TD-DOS, both in terms of reduction in cost-size and improvement in performances. In particular, the increase in area of new SiPMs detectors\cite{DiSieno2020Probe-hostedSpectroscopy} as well as the gating capabilities\cite{DallaMora2015FastOptics} means that larger photon travelling times and consequently penetration depths will be reached in the near future. What we still lack is a clear understanding of the physics and physiology of chest diffuse optics and further studies are needed to add additional insight and \textit{in vivo} validation.

\section*{Acknowledgments}
The authors acknowledge financial support from the European Union's Horizon 2020 programme under the BITMAP (n.675332) and LaserLab Europe (n.654148) grants. 

\section*{Disclosure}
The authors declare no conflicts of interest.
\bibliography{MendeleyPersonal}

\clearpage

\section{Supplementary Figures\label{SupplFig}}
The following figures will be attached as supplementary figures to the main figures, showing the other protocol (Prot5) with 5s duration of each phase and 10 repetitions (the first one with normal breathing), and the non-refolded analysis.

\setcounter{figure}{0}

\begin{figure}[ht!]
\centering\includegraphics[width=12.5cm, height=5cm]{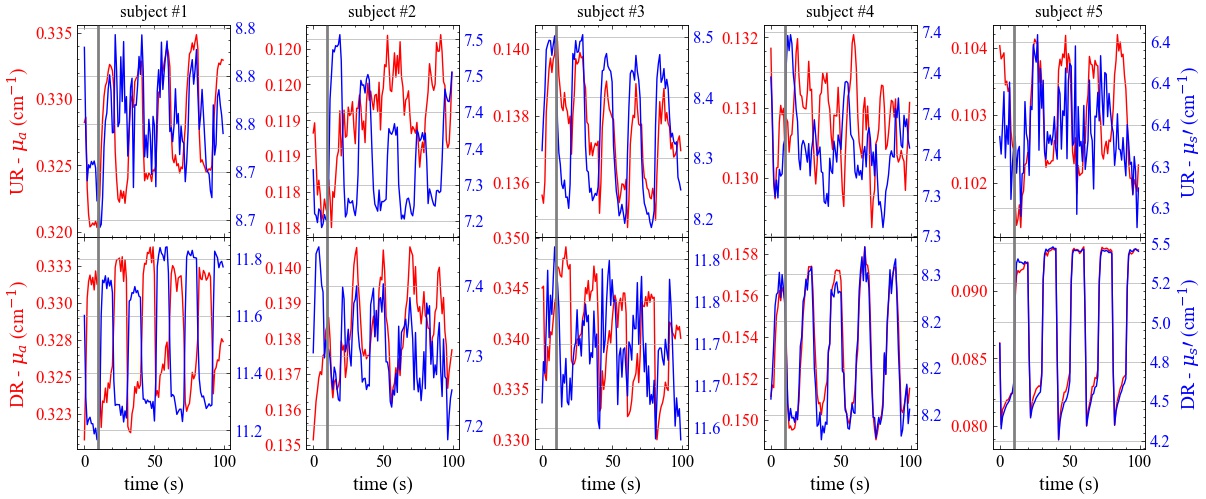}
\caption{Supplementary Figure~S\ref{FigOpt_b} to  Fig.~\ref{FigOpt}}
\label{FigOpt_b}
\end{figure}

\begin{figure}[ht!]
\centering\includegraphics[width=12.5cm, height=5cm]{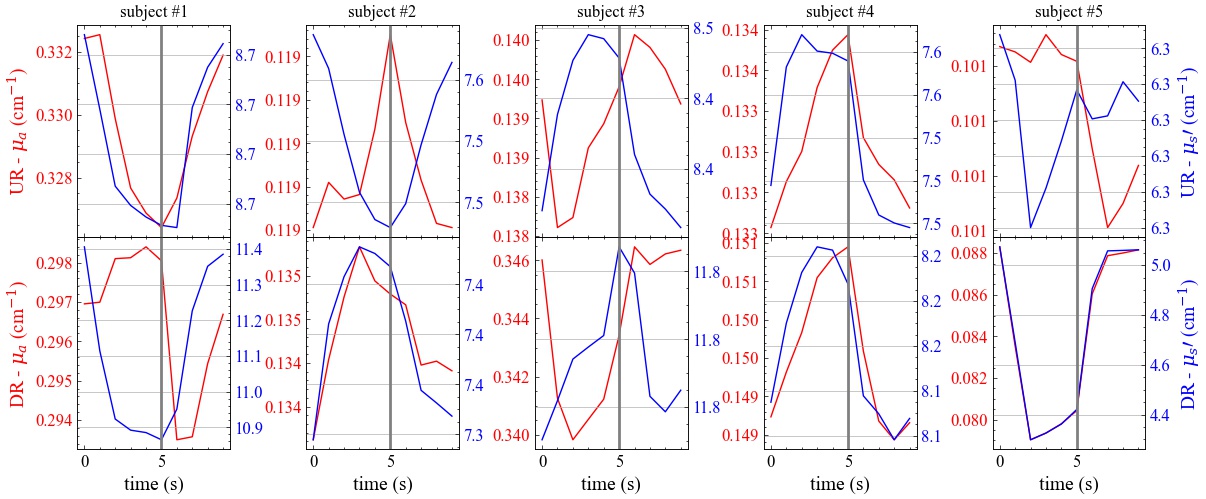}
\caption{Supplementary Figure~S\ref{FigOpt_c} to  Fig.~\ref{FigOpt}}
\label{FigOpt_c}
\end{figure}

\begin{figure}[ht!]
\centering\includegraphics[width=12.5cm, height=5cm]{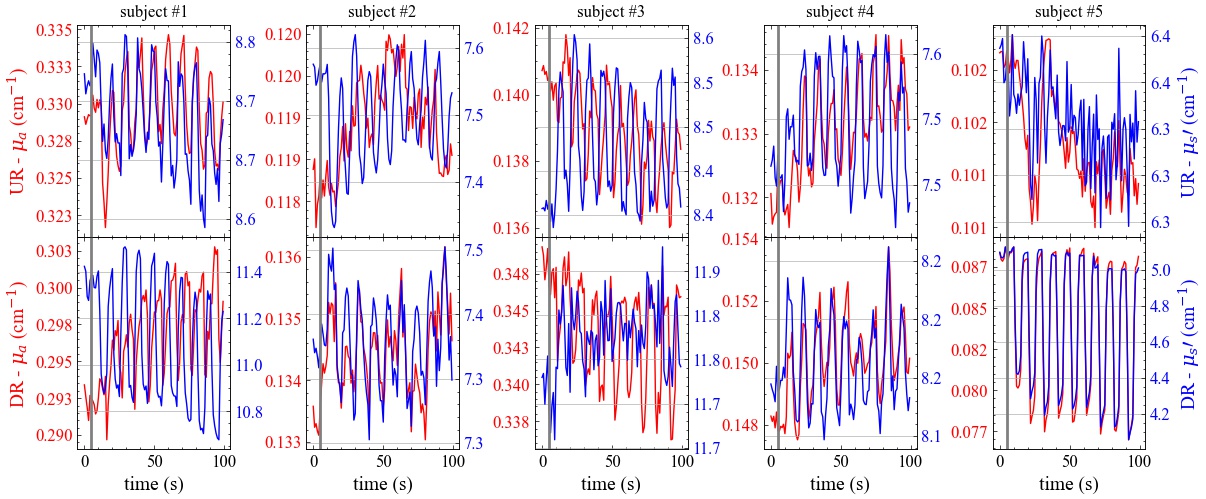}
\caption{Supplementary Figure~S\ref{FigOpt_d} to  Fig.~\ref{FigOpt}}
\label{FigOpt_d}
\end{figure}

\begin{figure}[ht!]
\centering\includegraphics[width=12.5cm, height=5cm]{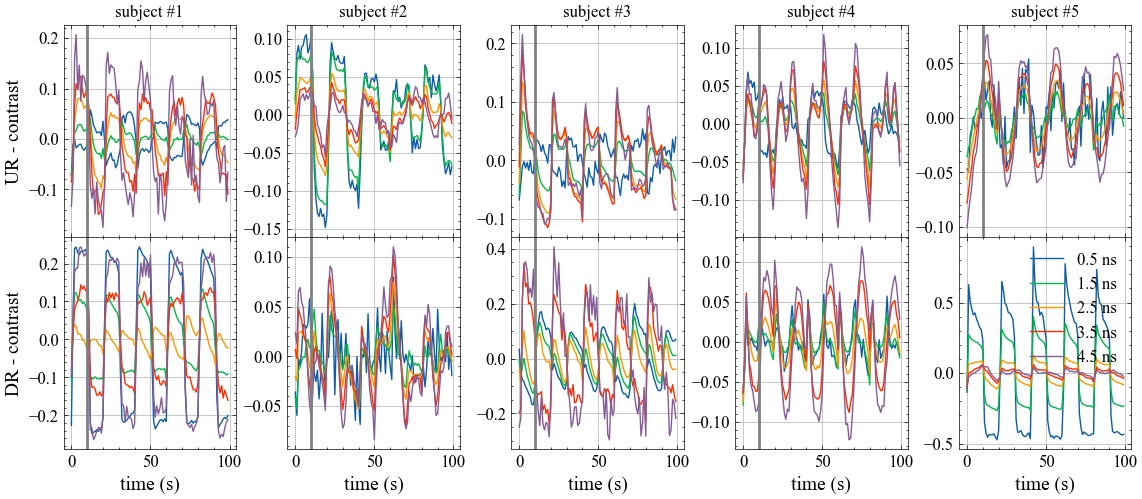}
\caption{Supplementary Figure~S\ref{FigGate_b} to  Fig.~\ref{FigGate}}
\label{FigGate_b}
\end{figure}

\begin{figure}[ht!]
\centering\includegraphics[width=12.5cm, height=5cm]{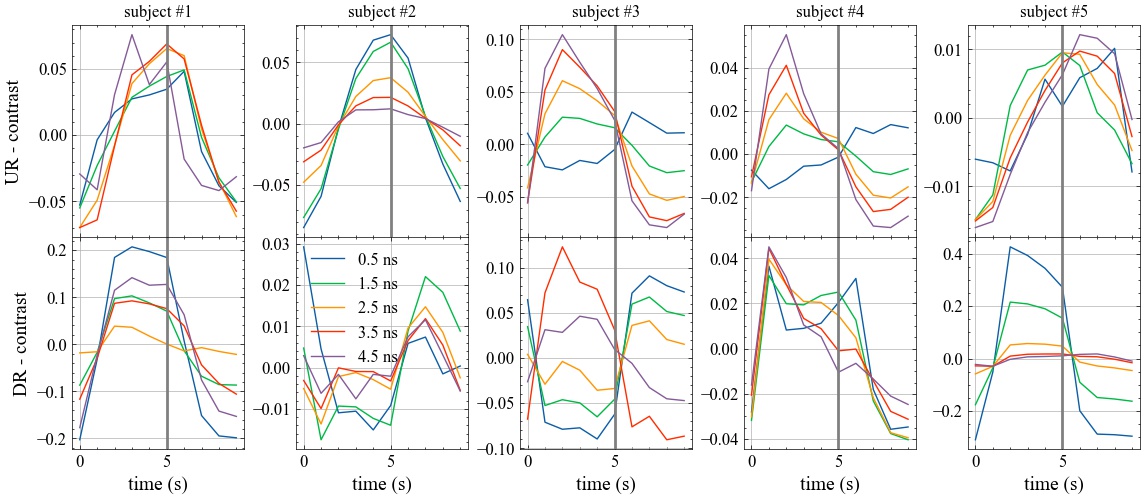}
\caption{Supplementary Figure~S\ref{FigGate_c} to  Fig.~\ref{FigGate}}
\label{FigGate_c}
\end{figure}

\begin{figure}[ht!]
\centering\includegraphics[width=12.5cm, height=5cm]{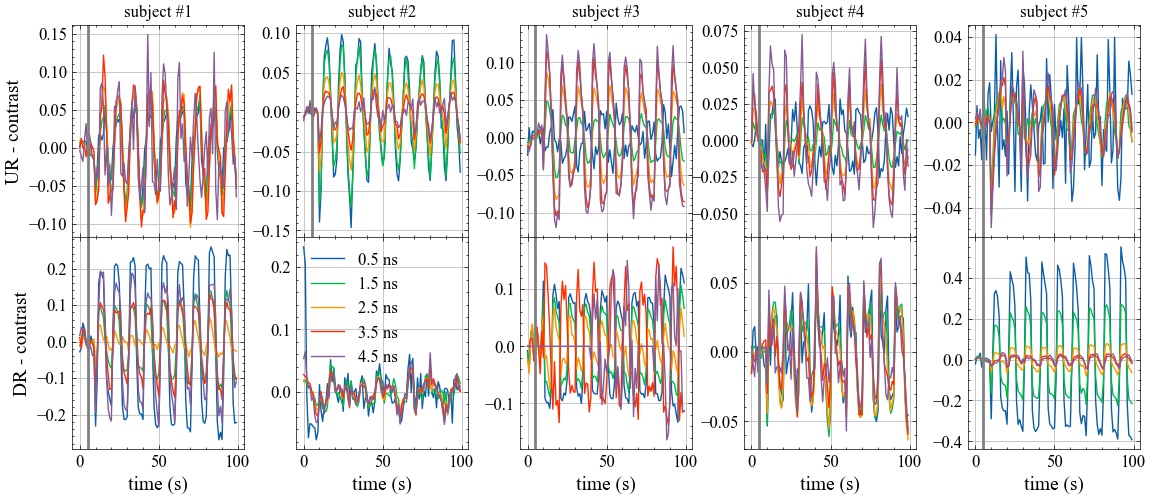}
\caption{Supplementary Figure~S\ref{FigGate_d} to  Fig.~\ref{FigGate}}
\label{FigGate_d}
\end{figure}

\begin{figure}[ht!]
\centering\includegraphics[width=12.5cm, height=5cm]{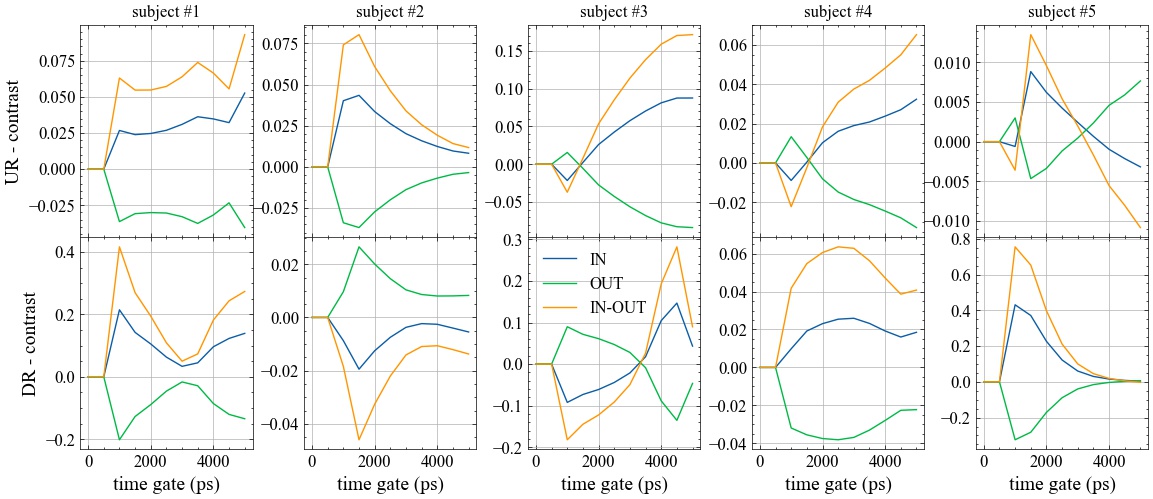}
\caption{Supplementary Figure~S\ref{FigMean_b} to  Fig.~\ref{FigMean}}
\label{FigMean_b}
\end{figure}

\end{document}